\documentstyle[preprint,prl,aps]{revtex}

\begin{document}

\draft    


\title{Absence of a pseudogap in the in-plane infrared response of 
Bi$_{2}$Sr$_2$CaCu$_2$O$_{8+\delta}$}
       
\author{A.F. Santander-Syro$^{1}$, R.P.S.M. Lobo$^{1}$, N. Bontemps$^{1}$, 
Z. Konstantinovic$^{2}$, Z.Z. Li$^{2}$ and H. Raffy$^{2}$}
\address{$^{1}$Laboratoire de Physique du Solide, 
         Ecole Sup\'erieure de Physique et Chimie 
         Industrielles de la Ville de Paris, CNRS UPR 5, 
         75231 Paris Cedex 5, France}
\address{$^{2}$Laboratoire de Physique des Solides, Universit\'e Paris-Sud
         91405 Orsay cedex, France}       
         
\date{\today}

\maketitle

\begin{abstract}
  The ab-plane reflectance of Bi$_{2}$Sr$_2$CaCu$_2$O$_{8+\delta}$ (Bi-2212)
  thin films was measured in the 30--25000~cm$^{-1}$ 
  range for one underdoped ($T_{c} = 70$ K), and one overdoped sample ($T_{c} = 63$ K)
  as a function of temperature (10--300 K). We find qualitatively similar behaviors 
  in the temperature dependence of the normal-state infrared response of both samples.
  Above $T_c$, the effective spectral weight, obtained from the integrated conductivity, 
  does not decrease when $T$ decreases, so that no opening of an optical pseudogap 
  is seen.  We suggest that these are consequences of the pseudogap opening first 
  in the ${\bf k}=(0, \pi)$ direction, according to ARPES, and of the in-plane infrared 
  conductivity being mostly sensitive to the $\bf {k}=(\pi, \pi)$ direction. 
\end{abstract}




The normal state of high-$T_c$ superconductors, in particular the pseudogap state, 
is highly debated \cite{tim1}. A number of experiments in underdoped samples 
(NMR \cite{alloul,ishida,williams}, specific heat \cite{loram}, 
uniform susceptibility \cite{watanabe}) show the opening of 
a gap in the density of states at some temperature $T^{\ast} > T_c$. 
Similarly, Raman spectroscopy points to a depletion of scattered intensity 
at low energy \cite{blumberg}. Nevertheless, the very existence of $T^{\ast}$ is 
disputed \cite{williams}. Tunneling data exhibit a broad dip above 
$T_c$ \cite{renner}, over the same energy range of the superconducting gap. 
ARPES measurements display a shift of the leading edge of the energy dispersion 
curve  spectrum in the ($0,\pi$) direction  \cite{ding}. They also show that, as 
temperature decreases, the opening 
of the pseudogap proceeds toward the  ($\pi,\pi$) direction, while keeping 
the shape of the d-wave superconducting gap\cite{norman}.

The in-plane DC resistivity in underdoped cuprates shows a deviation from the high 
temperature linear behavior. The temperature where this deviation occurs agrees 
with $T^{\ast}$ as determined by other techniques \cite{watanabe,chen}. In AC 
transport, the c-axis optical conductivity appears to track the temperature 
variation of the Knight shift \cite{homes}. The real part $\sigma_{1}(\omega, T)$ 
of the ab-plane infrared (IR) conductivity of underdoped samples exhibits  
a depression developing as $T(> T_c)$ decreases within an energy range extending 
typically from 200 to 1500 cm$^{-1}$ for the YBCO, Bi-2212 and Tl-2201 
compounds \cite{puchkov}. This loss of spectral weight (SW) was assigned either to the 
pseudogap opening or to a shift of SW towards very low frequencies \cite{puchkov}. 
It is obviously difficult to discriminate between these two approaches due to 
experimental limitations in the low  frequency part of the spectrum (usually 
restricted to $\omega \gtrsim 30$ cm$^{-1}$). 
A suppression of the optical  scattering 
rate $1/\tau(\omega)$ in underdoped (UD) samples below a typical constant energy scale 
$\sim 700$ cm$^{-1}$, irrespective of 
temperature and doping, 
was suggested to be the infrared signature of the pseudogap state \cite{puchkov,basov1,wang}.
None of these IR data  
convincingly defines a $T^{\ast}$ due to the small number of experimentally available 
temperatures \cite {Mihail}.

Indeed, to the best of our knowledge, no systematic study of the detailed temperature 
dependence of either the {\it in-plane} optical conductivity or scattering rate has 
been performed to date. In this work, we report a study of the in-plane infrared 
conductivity of Bi-2212 thin films for a set of 15 temperatures. Thanks to their 
highly reflective surface and their large area (typically $\geq 5 \times 5$~mm$^2$), 
thin films allow to resolve {\it reliably} relative changes in reflectance as small 
as 0.2 \%. This unprecedented accuracy allows us to look for the opening of a 
pseudogap by tracking a {\it transfer of spectral weight} from {\it low to high energies},
possibly occurring below $T^{\ast}>~T_c$.  Our data rule out such a direct 
signature of a pseudogap in the IR response, as they show the {\it opposite behavior}, 
namely spectral weight is collected at low energies as temperature decreases.

Thin films of Bi-2212 were epitaxially grown by RF magnetron sputtering onto SrTiO$_3$ 
substrates. DC transport in such films displays the features commonly assigned 
to the onset of a pseudo-gap. At $T_{c} < T < T^{\ast}$, the resistivity 
deviates from its linear temperature dependence, displaying a 
downward curvature in the UD samples. The overdoped (OD) samples show 
a slight upturn of the resistivity, hence no such $T^{\ast}$. The phase diagram 
thus derived was entirely compatible with the one obtained from e.g. ARPES
measurements in single crystals and similar films \cite{zorica,JCC}. Our spectra 
were recorded in (i) an underdoped film  (thickness 2400 \AA), 
with $T_{c}({\rm R} = 0) = 70 K$, and a relatively broad (onset-offset) 
transition width ($\sim 15 K$), and $T^{\ast} \sim 170 K$, according to the
resisitivity curve, and (ii) an  overdoped film (thickness 3000 \AA), 
with $T_{c}({\rm R}=0)=63 K$, and a transition width $\sim 5$ K.

Our data were collected with a Bruker IFS 66v interferometer (30 -- 7000 cm$^{-1}$). 
Near-infrared and visible data (4000 --
25000 cm$^{-1}$) were measured in a Cary 4000 grating spectrometer. In the 
overlapping spectral range, measurements agree within 1.5 \%. 
 The spectra  were 
measured for 15 temperatures between 300 K  and 10 K. 
 A He gas flow cryostat allows to stabilize the 
temperature within $\pm 0.2$ K. Figure~1 shows the  reflectivity
 of the two samples, for a restricted set of temperatures, up to 
1500~cm$^{-1}$.

Studying thin films precludes the use of the Kramers-Kronig (KK) transform to obtain the 
conductivity, due to the contribution of the substrate to the experimentally measured 
reflectivity. Therefore, in order to derive the optical conductivity of the films, 
we fit the reflectance of the film on top of a substrate using an attempt dielectric 
function for the film, and the optical constants of SrTiO$_3$ that we have experimentally 
determined for each temperature. The model dielectric function that fits the 
reflectance involves Lorentz and Drude oscillators, thus warranting 
causality. The procedure is described in detail elsewhere \cite{and1,and2}.
An interesting outcome of this approach 
is that it provides what can be considered as our best guess for the 
extrapolation of the dielectric function in the energy range which is 
not available experimentally, i.e. $\omega  \lesssim 30$~cm$^{-1}$. 

A systematic error in the absolute value of the reflectivity ($\pm 0.5 $~\%) 
is irrelevant since we are only interested in its relative changes with 
temperature. The change of a spectrum versus temperature is defined within 
$\Delta_{T} {\mathcal R} \leq 0.2$ \%, 
and the accuracy of the fit is $\Delta_{F} {\mathcal R} \leq 0.5$~\%. 
This  results into a typical error of 5 \% 
in the conductivity \cite{and3}. After completing 
the fitting procedure, we obtain all the relevant spectral functions for the film. 
We show in Fig.~2(a) the real part of the conductivity $\sigma_{1}(\omega, T)$ 
for the UD sample, at the same temperatures as in Fig.1. Figure~2(b) displays 
similar results for the OD sample. The DC resisitivity derived from our 
$\sigma_{1}(\omega, T)$ data are consistent with DC transport measurements 
on similar films. The so-called pseudogap effect shows up in the UD sample 
[Fig.~2(a)] as a depression in the conductivity above $T_c$, in the range 
$\sim 200- 1500$ cm$^{-1}$, as observed previously in single crystals. 
In the OD sample, a similar depletion of the conductivity occurs above 
$T_c$ in the same spectral range. The question is whether the observed loss 
is balanced by the increase of spectral weight at low energy. 

Before we address this point, we touch upon the scattering rate, defined as:
\begin{equation}
  \frac{1}{\tau(\omega)} =
     \frac{2\pi}{Z_0}\Omega^2_p {\rm Re}\bigg[\frac{1}{\sigma(\omega)}\bigg].
  \label{eq1}
\end{equation}
The plasma frequency $\Omega_p$ (in cm$^{-1}$) was obtained by integrating 
$\sigma_{1}(\omega)$
up to 1 eV.  $Z_{0} = 377 ~\Omega$ is 
the vacuum impedance.   Taking $\sigma(\omega)$ in $\Omega^{-1}$cm$^{-1}$, one gets  
$1 / \tau (\omega)$ 
in cm$^{-1}$. $1 / \tau (\omega)$ is shown in the insets of Fig.~2(a) and 2(b). 
As we are focusing on the normal state, we have only plotted $1 / \tau (\omega)$  
above $T_c$. In the UD sample, when the temperature decreases, a gradual depletion 
of $1 / \tau (\omega)$ below $\sim 700$ cm$^{-1}$ is observed [inset of Fig.~2(a)].
A similar, less pronounced, depletion below the same energy is also observed in the
OD sample already above $T_c$ [inset of Fig.~2(b)].  This fact can already be 
traced in those single crystals where doping was changed only by varying the amount 
of oxygen \cite{puchkov}, and has been ascribed in the underdoped regime to the opening 
of an optical pseudogap.

Our data show distinctly  the connection between the increase of the conductivity 
and the depression of the low energy scattering rate in {\it both samples}. 
The most striking result is the narrowing and increase of the low frequency 
Drude-like peak in the UD sample conductivity occuring {\it in the superconducting state},
with no obvious loss of spectral weight associated with the formation of the 
condensate. In contrast, in the OD sample, the conductivity drops below $T_c$, 
and the missing area is clearly visible. The superconducting state is discussed elsewhere 
\cite{and4}.

Since a set of 15 temperatures is available, we can look for a loss of spectral weight  
as the temperature is lowered. This would be the {\it actual signature of a pseudogap}
opening in the vicinity of the Fermi level. We have therefore integrated the 
conductivity in order to derive an effective carrier density up to $\omega_M$, 
according to:
\begin{equation}
  N_{eff} \propto
     \int_{0^+}^{\omega_{M}} \sigma_1(\omega)d\omega.
  \label{eq2}
\end{equation}

In order to display the thermal evolution starting from room temperature 
for various $\omega_M$ in a single plot, we have normalized 
$N_{eff}(\omega_{M}, T)$ with respect to $N_{eff}(\omega_{M},300$~K). 
We show in Fig.3, for both samples, the temperature variation of this 
normalized effective carrier density (equivalent to a normalized SW)
for a set of $\omega_M$ values  spanning the full experimental frequency 
range. When $\omega_M$ =100 cm$^{-1}$, the SW {\it increases} sharply as 
the temperature decreases down to $T_c$, showing that spectral weight is 
collected at low frequency in the normal state. As $\omega_M$  is increased, 
the transfer of SW becomes less apparent. The data for $\omega_{M}= 5000$ 
and $10000$ cm$^{-1}$ (not shown) can hardly be distinguished from those 
for $20000$ cm$^{-1}$. This shows that SW going to low frequency 
($\omega_{M} <100$ cm$^{-1})$ comes mostly from the mid-infrared range. 
The most important conclusion is that, for decreasing $T$, and
for all $\omega_M$ values considered (up to 20000 cm$^{-1}$), 
there is a systematic {\it increase of the spectral weight}, 
both in the UD and OD samples. 
This observation is at odds with the opening of a pseudogap which would 
yield a {\it decrease} of SW. Note that a break in this trend of increasing 
SW as $T$ decreases is seen only at the superconducting transition, where the 
SW hardly decreases in the UD sample, and drops sharply in the OD sample 
(Fig.~3) \cite{and4}.

So far, we have integrated Eq.~2 starting  at zero frequency, 
using the computed $\sigma_{1}(\omega)$ for $\omega < 30$ cm$^{-1}$. In order 
to check the sensitivity of our results to a low frequency cut-off, we have 
done a similar calculation, starting the integration from 30 and from 50 cm$^{-1}$. 
The result for the UD case is shown for $\omega_M = 1000$ cm$^{-1}$ in the inset of Fig.3. 
Introducing a low frequency cut-off results in a decrease 
of SW already at $T \lesssim 100$ K, and a broad maximum develops in temperature 
as the cut-off is increased. A similar behavior is observed for the overdoped 
sample. This being an artifact, confirms that no detectable loss of SW is seen, 
making it impossible to locate $T^{\ast}$ from the thermal evolution of the 
effective carrier density. Actually, from all the observations described above, 
it is  clear that there is enough  room in the low frequency part of the spectrum 
to balance the small depression of spectral weight observed in 
the conductivity data. Our data strongly suggest that the pseudogap has no 
clear-cut signature in the infrared conductivity.  Both the  depletion in the 
scattering rate and the occurence of a narrow peak in the real part of the 
conductivity are settling gradually as the temperature is decreased.

Since the existence of a pseudogap is established through a variety of 
experimental techniques, in both the spin and the charge channels, why 
does the optical conductivity appear not to display this phenomenon?
One logical hypothesis is that the conductivity probes mostly the quasiparticles 
(QP's) along the $(\pi ,\pi)$ direction, where no pseudogap opens 
\cite{ding,norman}. Indeed, recent ARPES experiments on $\it optimally$ 
doped Bi-2212 suggest that the in-plane transport might be dominated by nodal excitations 
\cite{JCCTau,VallaTau}, because single-particle scattering rates near the nodes and 
transport rates exhibit similar energy and temperature dependencies.   

One may explore how reasonable the above-mentioned hypothesis is in a simple framework:     
in the semiclassical approximation, the finite-frequency conductivity is \cite{Ashcroft}: 
\begin{equation}
\sigma \left( \omega \right) \propto \int \frac{{v}_{\mathbf{k}%
}^{2}}{\tau ^{-1}\left( \varepsilon _{\mathbf{k}}\right) -i\omega }\left( 
\frac{\partial f}{\partial \varepsilon}\right) _{\varepsilon
=\varepsilon _{\mathbf{k}}}d^{2}k\text{,}
\label{eq3}
\end{equation}  
where $\mathbf{v}_{\mathbf{k}}$ and $\varepsilon _{\mathbf{k}}$ are the
carrier bare (band-structural) velocity and energy, respectively. 
\ $\tau ^{-1}( \varepsilon _{{\bf k}}) $ is the transport scattering rate, 
and $f$ is the Fermi distribution function. Integration in Eq.~(\ref{eq3}) 
over the direction perpendicular to the Fermi surface (FS) shows that a critical 
weighting factor is the bare Fermi velocity ${\mathbf v}_{F}$, which 
---due to the anisotropy of the FS--- is smaller in the 
${\bf k}=(0,\pi)$ than in the ${\bf k}=(\pi ,\pi)$ direction. Simple calculations yield 
$v_{F}(\pi ,\pi)/v_{F}(0,\pi)\sim 5$ \cite{BokBouvier,Sherman2}. 
As a consequence, $\sigma_1(\omega)$
should be mostly sensitive to the nodal QP's, and no pseudogap
can be seen.  Furthermore, as 
the anisotropy of the FS holds in a sizeable doping range, the trends in the 
$T$-dependence of the normal-state $\sigma_1(\omega )$ should not strongly depend
on doping. This is indeed what our data show (Figs. \ref{fig2} and~\ref{fig3}).

Within this framework, the low-frequency narrowing of $\sigma_1(\omega)$
when temperature decreases can be assigned to the scattering of
nodal QP's (through coupling to low-energy excitations) becoming less efficient. 
An indirect contribution of the pseudogap may arise because 
less states are available for scattering in the ``pseudogapped'' regions, thus
enhancing the QP's relaxation time.  In the DC limit, this would
imply a downward deviation of the resistivity from its high-temperature behavior.

Up to now, ARPES measurements of the $\it renormalized$ Fermi velocity 
${\mathbf v}^{*}_{F}$ in optimallly doped Bi-2212 have reported an anisotropy of 
$v^{*}_{F}(\pi ,\pi)/v^{*}_{F}(0,\pi)\sim 3-4$ in the superconducting 
state \cite{VallaTau}.  Our results call for a deeper study 
of ${\mathbf v}^{*}_{F}$ as a function of doping and temperature, 
and of the relevance of the bare and renormalized Fermi velocities in the 
determination of the optical conductivity.

Among other interpretations, a gain in the low-frequency $\sigma_1(\omega )$ 
can be also regarded as a precursor of the coherent $\delta (\omega )$ peak
in the superconducting state, coming from either collective 
excitations \cite{Randeria}
or phase fluctuations of the superconducting gap \cite{EK}.
However, these models neglect the strong 
${\bf k}$-dependence of the electronic structure, of the gap and of the 
interactions in real materials. As our results demonstrate, these are 
crucial issues in the detailed understanding of transport phenomena 
in high-$T_c$ materials.

In summary, we showed that, at $T>T_c$, the in-plane IR response of overdoped 
and underdoped Bi-2212 are qualitatively the same. In particular, 
no loss of spectral weight, therefore no pseudogap signature, is seen 
in the normal state conductivity. We suggest that these are consequences  
of the anisotropy of the Fermi velocity.

The authors are grateful to E.~Andrei, J.~C.~Campuzano, R.~Combescot, G.~Deutscher, 
A.~Millis, M.~Randeria, E.~Ya.~Sherman and C.~Varma for illuminating comments. 
RPSML and AFSS acknowledge the financial support of CNRS-ESPCI, 
and Colciencias and the French Ministry of Foreign Affairs respectively.

\newpage

\begin{figure}
\caption{ Reflectance of (a) the underdoped and (b) the overdoped film.
 The sharp peaks  at 550, 120 and below 100~cm$^{-1}$, especially visible in the 
   underdoped sample, are the phonon modes of the substrate.}
\label{fig1}
\end{figure}

\begin{figure}
\caption{ Real part of the conductivity (from fit) of (a) the underdoped 
          and (b) the overdoped film. The insets show
          the scattering rate for $T>T_c$ (see text), using 
          $\Omega_{p}=16000$ cm$^{-1}$ and $16600$ cm$^{-1}$ respectively. 
          The line types refer to the same temperatures as in Fig. 1. The
          conductivity extrapolated from  30 cm$^{-1}$ down to zero 
          results from the fit (see text). }
\label{fig2}
\end{figure}

\begin{figure}
\caption{Spectral weight versus temperature for (a) the underdoped 
         and (b) the overdoped film.  Different symbols refer to 
         different cut-off frequencies $\omega_M$. 
         The inset shows the effect of 
         starting the integration from 0, 30 and 50~cm$^{-1}$ 
         (up to $\omega_M = 1000$ cm$^{-1}$).}
\label{fig3}
\end{figure}


\begin{references}
\bibitem{and2} As the film thickness is much smaller than the wavelength, 
we must consider coherent propagation of light in the film. 
We checked that our data are best described by assuming 
no reflection from the rear of the substrate, and multiple reflections 
within the film.
\end{references}
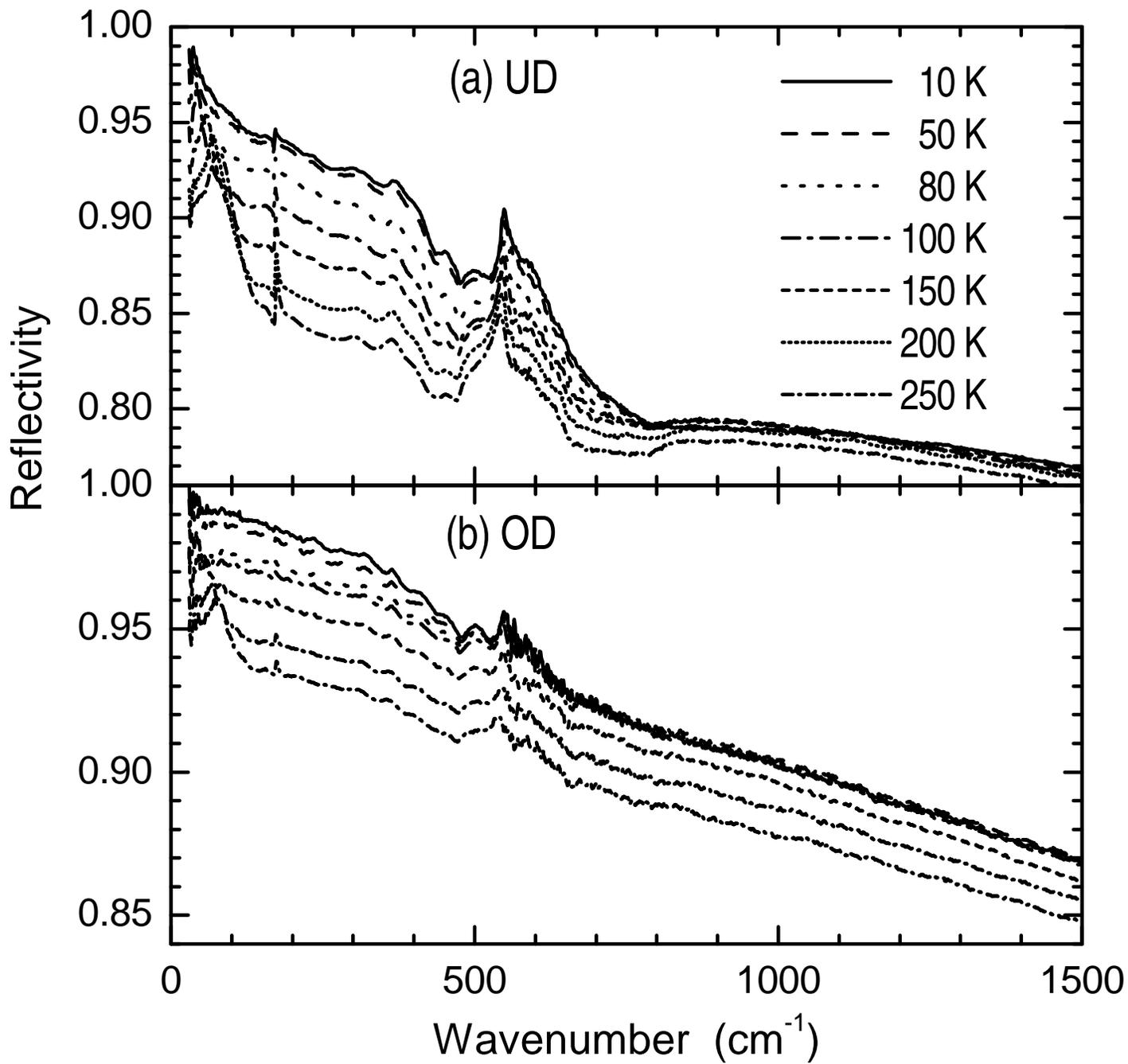
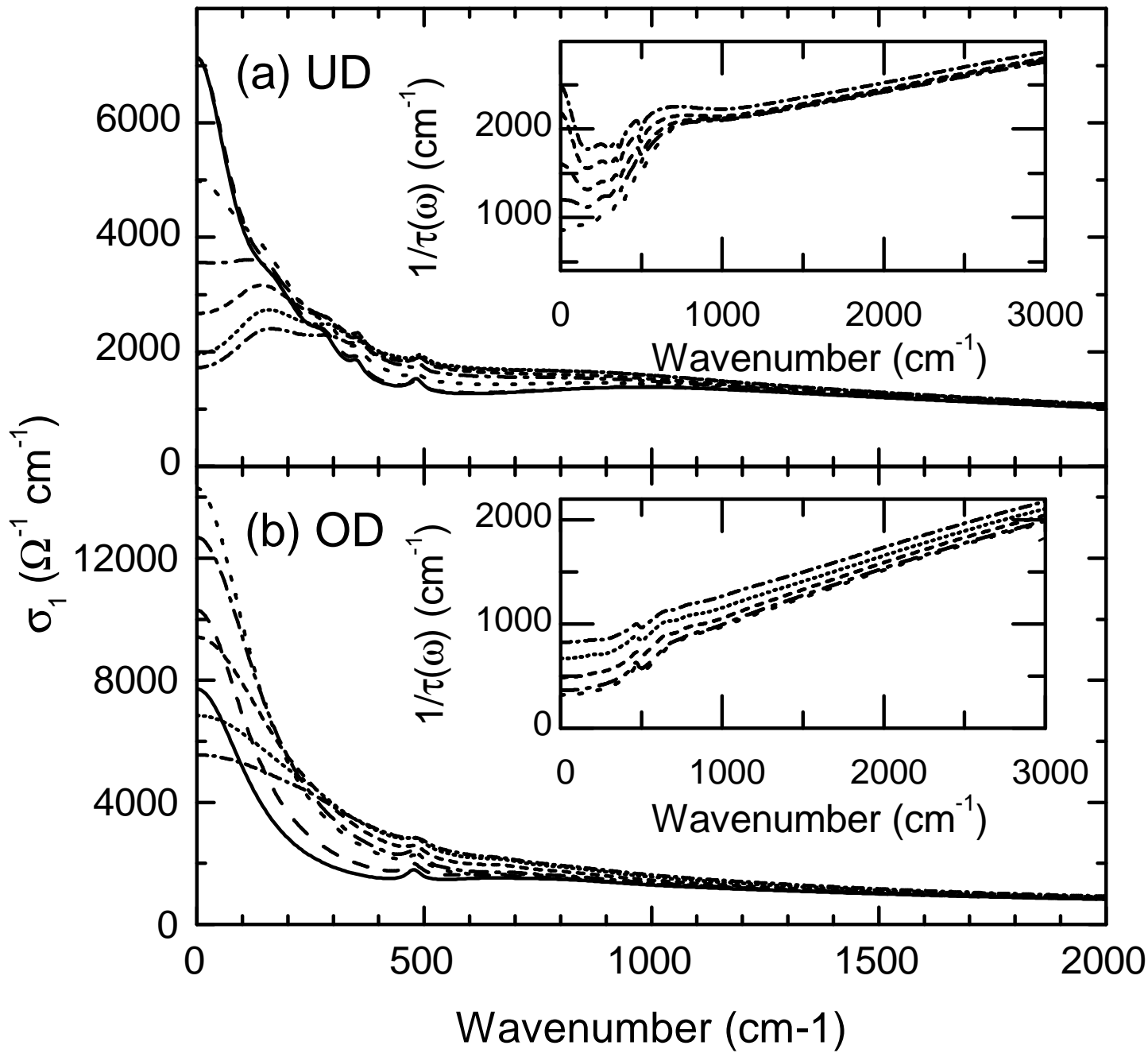
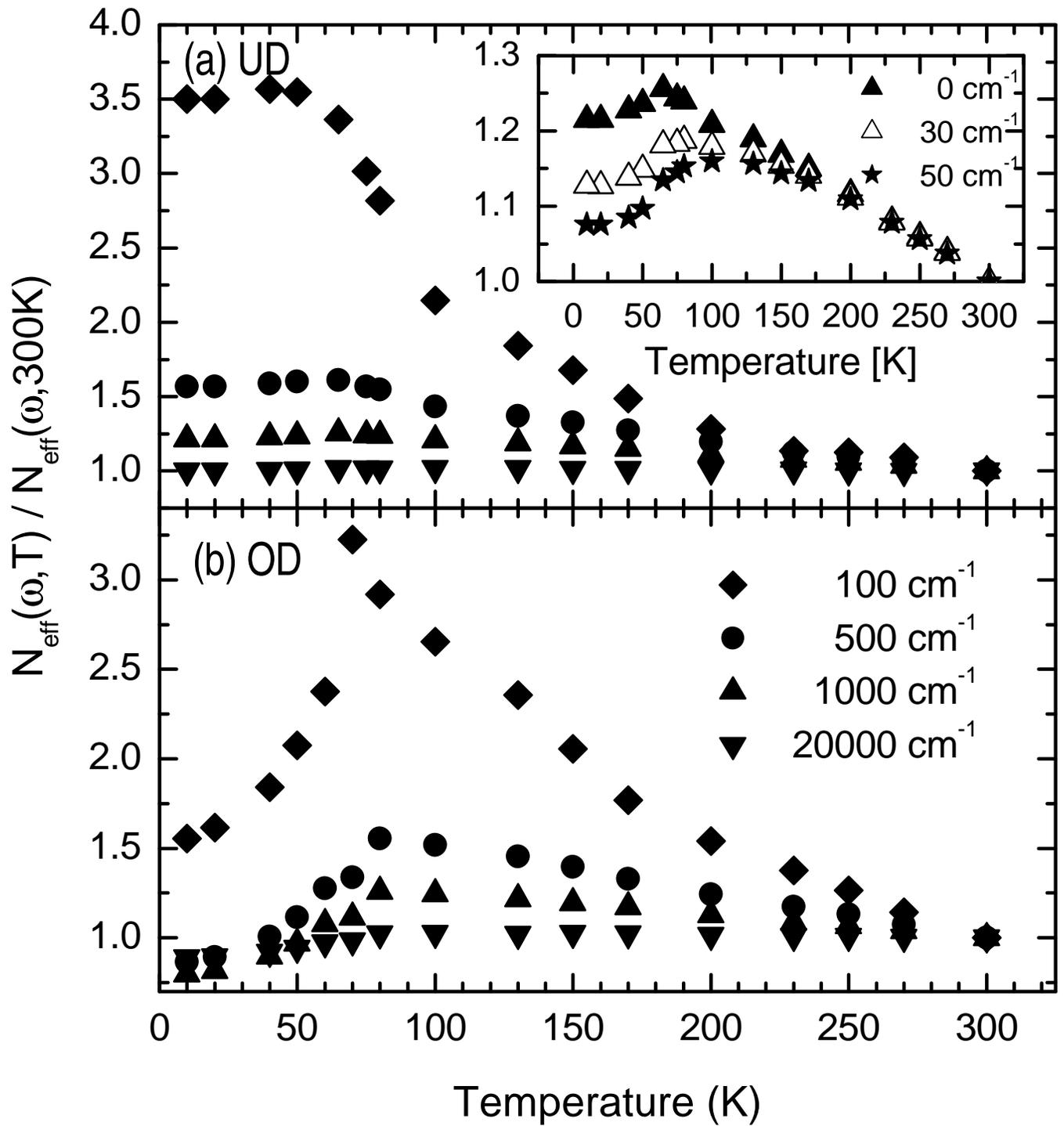
\end{document}